\documentclass[conference]{IEEEtran}
\IEEEoverridecommandlockouts

\usepackage{afterpage}
\usepackage{algorithmic}
\usepackage{array} %
\usepackage{amsmath,amssymb,amsfonts}
\usepackage{booktabs} %
\usepackage[nocompress]{cite}
\usepackage{color}
\usepackage{graphbox}

\usepackage{mathtools}
\usepackage{multirow}
\usepackage{textcomp}

\usepackage{tikz}
\usepackage{pifont} %
\usepackage{colortbl}
\usepackage{flushend}
\usepackage[normalem]{ulem}

\useunder{\uline}{\ul}{}
\usepackage{hyperref}
\usepackage{cleveref} 
\usepackage{atbegshi} 
\usepackage[disable]{todonotes}
\usepackage{mfirstuc}
\usepackage{csquotes}

\usepackage{textcomp}
\usepackage{wrapfig}

\newcommand{\name}{CLUE} %
\newcommand{\Name}{\xcapitalisewords{\name{}}} %
\newcommand{\NameLong}{(Cloud-native Sustainability Evaluator)}

\newcommand{\anonymize}[1]{#1}

\newcommand*\circled[1]{
    \hspace{-0.5em}
    \tikz[baseline=(char.base)]{
            \node[shape=circle,draw,inner sep=0.75pt] (char) {#1};
    }
    \hspace{-0.5em}
}

\newcommand{\comp}[1]{\textbf{#1}}
\newcommand{\metric}[1]{\textbf{#1}}
\newcommand{\high}[1]{\textcolor{red!80}{\textbf{#1}}}
\newcommand{\low}[1]{\textcolor{teal!80}{\underline{\textbf{#1}}}}

\newcommand{\todocite}[1]{}

\def\BibTeX{{\rm B\kern-.05em{\sc i\kern-.025em b}\kern-.08em
    T\kern-.1667em\lower.7ex\hbox{E}\kern-.125emX}}
\begin{document}

\makeatletter %
\newcommand{\linebreakand}{%
  \end{@IEEEauthorhalign}
  \hfill\mbox{}\par
  \mbox{}\hfill\begin{@IEEEauthorhalign}
}
\makeatother %

\newcommand\submittedtext{%
\footnotesize \textbf{Preprint}. This work has been accepted to the 22nd IEEE International Conference on Software Architecture (ICSA'25). Copyright may be transferred without notice, after which this version may no longer be accessible.}

\newcommand\submittednotice{%
\begin{tikzpicture}[remember picture,overlay]
\node[anchor=south,yshift=10pt] at (current page.south) {\fbox{\parbox{\dimexpr0.65\textwidth-\fboxsep-\fboxrule\relax}{\submittedtext}}};
\end{tikzpicture}%
}

\title{A Comprehensive Experimentation Framework for Energy-Efficient Design of Cloud-Native Applications}

\author{\IEEEauthorblockN{Sebastian Werner}
\IEEEauthorblockA{\textit{Information Systems Engineering} \\
\textit{Technische Universität Berlin}\\
Berlin, Germany \\
sw@ise.tu-berlin.de}
\and
\IEEEauthorblockN{Maria C. Borges}
\IEEEauthorblockA{\textit{Information Systems Engineering} \\
\textit{Technische Universität Berlin}\\
Berlin, Germany \\
mb@ise.tu-berlin.de}

\linebreakand

\IEEEauthorblockN{Karl Wolf}
\IEEEauthorblockA{\textit{Information Systems Engineering} \\
\textit{Technische Universität Berlin}\\
Berlin, Germany \\
kw@ise.tu-berlin.de}
\and
\IEEEauthorblockN{Stefan Tai}
\IEEEauthorblockA{\textit{Information Systems Engineering} \\
\textit{Technische Universität Berlin}\\
Berlin, Germany \\
st@ise.tu-berlin.de}
}

\maketitle
\submittednotice

\begin{abstract}
Current approaches to designing energy-efficient applications typically rely on measuring individual components using readily available local metrics, like CPU utilization. However, these metrics fall short when applied to cloud-native applications, which operate within the multi-tenant, shared environments of distributed cloud providers. Assessing and optimizing the energy efficiency of cloud-native applications requires consideration of the complex, layered nature of modern cloud stacks.

To address this need, we present a comprehensive, automated, and extensible experimentation framework that enables developers to measure energy efficiency across all relevant layers of a cloud-based application and evaluate associated quality trade-offs. Our framework integrates a suite of service quality and sustainability metrics, providing compatibility with any Kubernetes-based application. We demonstrate the feasibility and effectiveness of this approach through initial experimental results, comparing architectural design alternatives for a widely used open-source cloud-native application.   

\end{abstract}

\begin{IEEEkeywords}
Cloud-Computing, Sustainable and Quality Engineering, Experiment-driven Software Design
\end{IEEEkeywords}

\section{Introduction}\label{ch:intro}

Cloud computing promises virtually limitless resources, fueling a wave of new cloud-native applications designed to leverage this potential. Yet, as cloud usage soars, so does its energy demand, putting a strain on our planet's finite resources. Cloud data centers already account for approximately 3\% of the global energy consumption \cite{knowles_datacenternumbers_2021}, with demand expected to continue growing over the next decade \cite{Andrae_datacentertrends2023_2020}. To combat this worrying trend, cloud providers and application developers must adopt more energy-efficient practices and implement strategies and software architectures that minimize the carbon footprint. %
Even though most cloud providers have already taken measures to lower their carbon emissions \cite{pedram2012efficientdatacenters}, little has been done to address emissions and energy consumption from the application side \cite{vitali2022towards,voslago_energytacticscloud_2022}. 

Cloud-native applications are typically designed to be available, scalable, and resilient \cite{leitner2015cloudnativems}. Designing for these qualities often involves trade-offs that can compromise energy-efficiency. Further, the agile nature of modern application development encourages the rapid development of new services, and, in the rush to innovate, energy-efficiency can frequently be overlooked. %
As sustainability becomes an increasingly important software quality \cite{kazman_energyefficiencyinarchitecture_2018}, this will need to change.

Recent research has started to address this challenge by integrating energy efficiency considerations into the architectural design process \cite{kazman_energyefficiencyinarchitecture_2018}, and by proposing several architectural tactics  \cite{procacciantiLago_greentactics_2014,paradiskazman_energytacticsreview_2021,voslago_energytacticscloud_2022, martinezLago_energytacticsreview_2023} to manage energy efficiency. However, to assess the impact of such tactics and weigh between design alternatives, application developers need appropriate methods and tools that can measure the energy efficiency of these tactics in their application and reveal energy-related trade-offs.

 Consequently, the need for such tools has been recognized by research \cite{paradiskazman_energytacticsreview_2021,amaral_kepler_2023} and industry\footnote{\url{https://greensoftware.foundation}} \footnote{\url{https://www.cloudcarbonfootprint.org}} alike.  
However, unlike qualities like cost, energy efficiency and carbon footprint are less accessible to developers, as tooling and assessment methods are only beginning to emerge \cite{lennick_microservice-based_2021,vitali2023enriching,wiesner2021wait,schirmer2023night,stojkovic2023ecofaas}. 
 Measuring the energy efficiency of cloud-native applications is particularly challenging. These applications are built as compositions of layered, distributed, and diverse services, deployed as containers or serverless functions, and rely on shared cloud infrastructure.  Existing experiments often focus solely on CPU utilization or individual components \cite{dinga2023empirical}, failing to address these complexities. 
Similarly, other approaches, such as in \cite{funkelago_energytacticsexperiments_2024}, rely on estimates and omit virtualization abstractions. However, it is important to consider each layer and deployment tool, as they introduce overhead and create potential for inefficiencies.

In light of these complexities, we argue that a comprehensive measurement and assessment approach is needed. To that end, we present \name{}, a comparative experimental benchmarking framework to assess the energy-efficiency of cloud-native applications. It incorporates application-centric service quality and sustainability
metrics in an extensible and easy-to-use tool, allowing developers to understand energy-related trade-offs and weigh between different architectural design alternatives in their applications. 

Towards this end, we present the following contributions:
\begin{enumerate}
    \item An experiment design and metrics for evaluating the energy-efficiency of cloud-native architectures.
    \item \Name{} - an automated experimentation framework that applies the experiment design.
    \item We evaluate the framework on a well-known cloud-native application \cite{eismann_teastore_2019}, and show its use in assessing the energy efficiency of different design alternatives.
\end{enumerate}
The remainder of the paper is structured as follows: we review relevant background and related work (sec. \ref{ch:bgrw}), present our method (sec.
\ref{ch:design}) and tool (sec \ref{ch:main}), showcase the applicability of the framework for evaluating trade-offs in an open source application (sec. \ref{ch:eval}) and conclude in sec. \ref{ch:conclu}.

\begin{figure*}[ht!]
    \centering
    \includegraphics[width=0.7\textwidth]{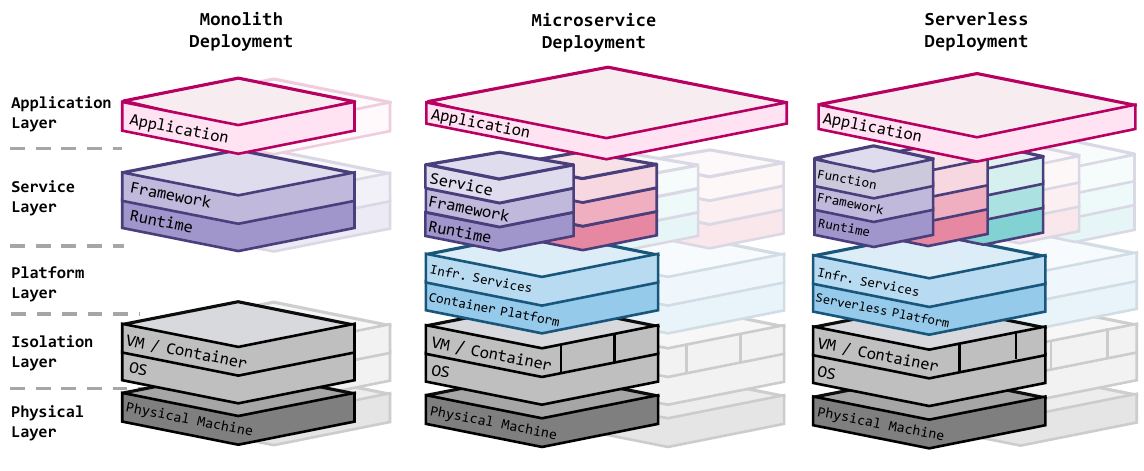}
    \caption{The multi-layered nature of modern cloud-native software systems}
    \label{fig:layers}
\end{figure*}
\section{Background and Related Work}\label{ch:bgrw}

In this section, we provide background on cloud-native applications and discuss related work.

\subsection{Cloud-native Applications}

Cloud applications can be deployed using different approaches, and leveraging many technologies across the stack. Figure \ref{fig:layers} depicts the cloud software stack. 
Traditionally, they are implemented as monolithic applications running on virtual machines (VMs). Here, engineers typically control the guest operating system, where they install the necessary runtimes (e.g., the Java Runtime) and frameworks (e.g., Hibernate ORM) required to run their monolithic applications.

More modern applications are cloud-native, i.e. tailored to take full advantage of the cloud by leveraging the microservice or serverless architecture paradigms. 
In microservice architectures, the application is broken down into smaller, independently deployable services. Each microservice runs in its own container, and is orchestrated by a container platform (e.g., Kubernetes). This abstraction layer can be complemented by other infrastructure services, like sidecars and service meshes. Microservices enable developers to use different runtimes and frameworks for each component of the application, and can be scaled independently.

In serverless architectures, the unit of execution is even smaller, typically a function rather than a whole service. Developers can choose from various runtimes provided by the cloud provider, and while frameworks can be used, they should be lightweight to maintain performance. Serverless functions run on top of containers or microVMs, depending on the provider, and are managed alongside the runtime by the platform provider. Thus, they reduce the operational task of the developer even further than container orchestration platforms.

\subsection{Architecting for Energy Efficiency}
As illustrated in figure \ref{fig:layers}, cloud-native computing introduces a large and complex set of abstractions throughout the stack. The same abstractions that provide benefits in scalability and fast-paced development can also introduce overhead and impact energy efficiency. At the same time, the cloud also provides many possibilities for efficiency gains, e.g., through pooled resources \cite{procacciantiLago_greentactics_2014} or through on-demand scaling \cite{voslago_energytacticscloud_2022}.  From a practitioner’s perspective, there are many ways to design an application, making it essential to experiment with different design alternatives to identify more efficient solutions. Architectural tactics focused on energy efficiency provide a blueprint for this exploration. %

The work of Procaccianti et al. \cite{procacciantiLago_greentactics_2014} first introduced architectural tactics for energy efficiency. This work is then extended by Paridis et al. \cite{paradiskazman_energytacticsreview_2021} and Vos et al. \cite{voslago_energytacticscloud_2022}, with \cite{martinezLago_energytacticsreview_2023} providing a current review. The review collected over 160 architectural tactics. Here, we highlight a few relevant to the cloud-native application stack.

In the service layer, developers can employ tactics like \textit{T: Increase Efficiency} \cite{paradiskazman_energytacticsreview_2021}, which involves making use of efficient data structures and algorithms in their code, or \textit{T: Adopt Use-case Driven Design} \cite{voslago_energytacticscloud_2022}, which suggests refactoring services that don't contribute to business value to be more resource economical. Further, they can reduce or replace framework dependencies, as these also play an important role in the energy efficiency of the application \cite{ProcacciantiLago_ORMenergyconsumption_2016}. In the runtime, developers can \textit{T: Choose an Efficient Runtime}. Programming languages have varying levels of overhead, with some being more resource-intensive than others, as investigated by \cite{pereira_languagesoverhead_2017}. Even different versions of the same runtime can impact energy consumption in distinct ways, as highlighted in \cite{Ournani_JVMenergyconsumtion_2021}.

The platform layer also provides numerous opportunities for exploration. In line with expectations, adding container abstraction and management layers introduces overhead, as demonstrated by \cite{santos_dockeroverhead_2018} for Docker, \cite{truyen_k8soverhead_2019} for Kubernetes, and \cite{elkhatib_istiooverhead_2023} for Istio. The tactic \textit{T: Reduce Overhead by Removing Intermediaries and Abstractions} \cite{paradiskazman_energytacticsreview_2021} could be implemented by changing the deployment paradigm to a monolith, hereby removing these abstractions. Here, Paradis et al. \cite{paradiskazman_energytacticsreview_2021} predict an improvement in overhead, at the cost of modularity.
On the other hand, efficiency improvements on the platform layer could also be achieved through a serverless deployment. The tactics \textit{T: Apply Granular Scaling} and \textit{T: Choose a Fitting Deployment Paradigm} \cite{voslago_energytacticscloud_2022} encourage the switch to serverless for services with infrequent, bursty workloads, because the serverless platform is supposed to be able to scale from zero on demand.

This represents only a subset of possible tactics, yet it already provides ample room for exploration. To facilitate this exploration, architects must be equipped with appropriate measurement and experimentation methods.

\subsection{Measuring and Managing Energy Efficiency}
Cloud vendors currently do not equip users with the necessary transparency and tools to effectively measure the energy efficiency of their applications\cite{mytton2020assessing,voslago_energytacticscloud_2022}. The provided dashboards offer only a very coarse view on energy consumption. Google's Dashboard, for example, only provides monthly consumption averages and does not allow to drill down into the consumption of individual services. 
Since the support from cloud vendors is lagging behind, many researchers have started to investigate alternative approaches to estimate the energy efficiency of cloud-native applications.

Here, Lannelongue et al.~\cite{lannelongue2021green} for example introduce a methodological framework and an online tool, Green Algorithms, to estimate the carbon footprint of computational tasks, based on meta-data about the hardware and runtime used.
We also see similar works in the context of evaluating Kubernetes-based microservice environments in \cite{centofanti2024impact}. The authors investigate the accuracy and reliability of existing open source energy measuring tools such as Scaphandre and Kepler \cite{amaral_kepler_2023}. The tools still show some discrepancies to hardware sensors, a limitation that we take into account in our work. 

Besides the current lack of very accurate measurement tools, there is already a considerable effort to explore and evaluate the impact of different layers in cloud-native applications (see \Cref{fig:layers}).
For example for the platform layer, Sharma et al.~\cite{sharma2023challenges} explore energy inefficiencies of serverless computing, highlighting how the inherent attributes of serverless functions lead to significantly higher energy consumption compared to conventional web services. 
Similarly, Lennick et al.~\cite{lennick_microservice-based_2021} %
present a performance and energy analysis of a microservice architecture for IoT systems using Docker containers.
Other approaches explore the impact of infrastructure services, such as Dinga et al.~\cite{dinga2023empirical} which evaluated the energy efficiency of different observability platforms. 

Researchers have also proposed concrete approaches to reduce energy efficiency on an application and service level. %
The approach by Wiesner et al.~\cite{wiesner2021wait}, which examines the potential for reducing data center emissions by shifting computational workloads to times when energy supply is less carbon-intensive, focusing on regional differences and temporal workload shifting. This approach is further explored by Schirmer et al. \cite{schirmer2023night}, and applied specifically to serverless applications.
Lastly, the work of Vitali~\cite{vitali2022towards} rethinks application design for energy efficiency, similar to the tactic of use-case driven design described in the previous section. The author proposes reducing the quality of certain services to a more energy-efficient variant when there is a need for the platform to save energy. 

We see such approaches as complementary to our work, as they help practitioners come up with new tactics that they can then assess in the context of the entire application.
With our work, we aim to provide practitioners with an extensible tool that can be used to evaluate these different tactics, technologies and measurement tools in the context of their unique combination of abstractions and technology choices. Assessments should be done for the entire application, instead of using focused micro-benchmarks or theoretical settings, enabling developers to start building leaner applications today.

\section{Methodology}\label{ch:design}

In this paper, we focus on the quality trade-off between energy efficiency and service quality. We argue that effective and developer-oriented measurements of sustainability in cloud-native applications need to be architecture-centric and reflect changes in service quality, and energy efficiency in the context of regular use. 
Thus, we follow an experiment-based methodology~\cite{boreges_observe_icsa_2024} and in the following, introduce an experiment design targeted to measure these qualities during development. %

\subsection{Metrics}
At the core of the experiment design, we propose a set of metrics that both address the different layers of the application architecture as depicted in \Cref{fig:layers} and at the same time allow for sufficient abstraction to account for the complex interdependencies of modern cloud-native applications.
We are not necessarily interested in the absolute values of these metrics but in the relative changes between application modifications. 
Current tooling is often not accurate when it comes to measuring energy consumption, as for example, thermal loads are also necessary to derive the emissions of the hardware used. 
Hence, we work with what is available, but also assume that with the increasing demand for sustainability measurements, more accurate tools will be developed, and that cloud providers will start to offer finer granularity for emission-related metrics.

\begin{table}[]
  \centering
  \caption{The core metrics used to evaluate the sustainability and quality of an application.}
  \label{tab:metrics}
  \resizebox{\columnwidth}{!}{
  \renewcommand{\arraystretch}{1.3}

\begin{tabular}{llp{.65\columnwidth}}
  \textbf{Name}                & \textbf{Unit} & \textbf{Description} \\ \hline
  \multicolumn{3}{c}{\textbf{Sustainability Metrics}} \\ \hline
  Request Consumption ($WR$) & $Ws$              & Energy used per average request, excluding platform and isolation layers. \\
  Runtime Overhead ($RO$)    & $[0..1]$      & A percentage of the overhead of the consumed resources (incl. platform and isolation layers) relative  to the application's code. \\
  Resource Utilization ($RU$) & $[0..1]$         & The amount of resources used against the total amount of resources provisioned. \\ 
  Resource Efficiency ($RE$) & $Ws$      & The energy that is wasted due to inefficient (delayed) scaling of resources to match the demand.\\
  Auxiliary Costs ($AC$) & $Ws$ & Additional consumption due to network, storage, and cooling in the cloud environment.\\
    \hline
  \multicolumn{3}{c}{\textbf{Quality Metrics}}                                                                    \\ \hline
  Total Costs ($TC$)        & $\$$               & Normalized cost for this deployment, assuming second-based billing. \\
  Failure Rate ($FR$)        & $[0..1]$          & The amount of request failures against the total amount of requests. \\
  \hline
  \multicolumn{3}{c}{\textbf{Performance Metrics}}                                                                \\ \hline
  Requests ($Rqs$)             & $\mathbb{R}$    & The requests per second during normal operation. \\
  Latency ($Lat$)              & $s$           &  End-to-end latency of requests. \\ \hline
\end{tabular}

  }
\end{table}

We introduce 9 initial metrics (see \Cref{tab:metrics}), grouped into three categories: sustainability, application quality, and application performance. 
We define these metrics implementation independently, as current measurement methods and tools are expected to improve over time, e.g., we might be able to read the water use of cloud instances and thus obtain sustainability measurements beyond electricity.

A key sustainability metric is $WR$, which aims to assess the \metric{mean consumption per request} during operation, and shall, consider the consumption of every aspect specifically involved in handling the application's requests.
For example, this would include the Java Runtime, Application Server (e.g., Tomcat), and the business logic for a Java-based web service, but exclude the consumption of the Kubernetes platform, as these emissions are shared across all services or even multiple users and need to be accounted for differently.
Second, we consider the \metric{runtime overhead} ($RO$), which is all the overhead (in consumption, cooling capacity, ...) that arises from the provision of resources to these services. 
For example, here, we would measure the consumption required to run Kubernetes or to offer Serverless functionality while excluding resource usage caused by the service resources themselves.
While these overheads are typically shared between applications in cloud environments, the impact factor of a selected technology should nevertheless be considered when making development choices.
If the carbon intensity of the electricity used by that data center and its hardware is known, this metric allows computing the Software Carbon Intensity (SCI) of the application variant under test.
Third, we consider \metric{resource utilization} ($RU$), which reflects the amount of resources used during measurement in comparison to the total amount of provisioned resources, i.e., based on the choice of server, resource limits, and specifically scaling decisions. 
\metric{Resource efficiency} ($RE$) aims  to measure how well an application manages to
utilize the available resources to provide the desired service quality without under- or over-provisioning.
Note that $RO, RU$ are typically directly reported in cloud environments, due to the pay-per-use model. 
Hence, one way to calculate these metrics across a diverse set of provisioned resources is to use the cost of the particular cloud provider.
Lastly, we consider \metric{auxiliary hosting costs} ($AC$), which occur for the cloud environment to provide this application, such as networking, storage, and cooling of auxiliary equipment such as network switches. 
This metric is certainly hard to measure, in a dynamic and vastly distributed system such as modern cloud infrastructure, however, estimation methods exist\footnote{\url{https://sci.greensoftware.foundation/}}.

For general application quality, we consider the \metric{total monetary cost} ($TC$), i.e., the cost required to provision the application for the given cloud provider. With this, we aim to consider and detect trade-offs in cost efficiency. For on-premises cloud solutions, a reference cost model, e.g., costs of equivalent services AWS could be used to get an estimate here. 
Additionally, we track the \metric{failure rate} ($FR$) as a percentage of failed requests that are processed during the measurement period to account for possible reliability trade-offs. 
Finally, we track common performance metrics, i.e., the number of \metric{requests per second} ($Rqs$) and mean end-to-end \metric{latency} ($Lat$) that a deployment is delivering.

For all these metrics, we stipulate that for a fair comparison, the same measurement methods need to be used, and thus, if one approach currently offers only limited energy, hardware, or emissions observability, we must accept overall reduced quality and use estimation models.
Moreover, in order to mitigate measuring artifacts from operating on a single layer, we recommend measuring throughout the application stack.
In \cref{sec:evalprot}, we present how we implemented these metrics in the case of \Name{} (see also \cref{sec:clue}).

\subsection{Design}

With the proposed metrics, we can now characterize the sustainability and quality of a change in the application, provided we observe the application before and after the change for a sufficiently long time.
For this, we require a system that can be easily deployed or observed in both states.
Hence, we assume a setting, that allows the creation of the entire application programmatically, e.g., through infrastructure as code, into a test or staging environment.

Moreover, the experiment design requires a representative use of the application in the form of one or more artificial workloads or by collecting sufficient production observations. 
We assume that a developer can provide realistic and representative artificial workloads to enable the collection of these metrics.
However, one aspect that should also be observed specifically for energy consumption is the idle state of a system, as this already gives an indication as to how much overhead is introduced due to providing all necessary components.

Lastly, we assume that the application can be observed in sufficient granularity and that we can observe as many of the application layers as possible. 
Hence, the underlying infrastructure and technologies should offer necessary instrumentation and observation points.

These aspects, however, are all common in cloud-native applications, which often due to complex configurations already rely on infrastructure as code,  use artificial workloads for integration tests in staging environments, and use a growing stack of observability tools for reliability engineering.
Thus, applying the methodology for evaluating the application of specific sustainability tactics should be possible with minimal additional effort given the right tooling support.

\section{\Name{}}\label{ch:main}\label{sec:clue}
Implementing the experiment design, we created \name{} \NameLong{} -- an experimentation and observability framework to gather and compile the sustainability and quality changes in a cloud-native application, an overview of which can be seen in \Cref{fig:exparch2}. \Name{} is implemented in Python and available on GitHub\footnote{\anonymize{\url{https://github.com/ISE-TU-Berlin/Clue}}}.

\subsection{Components}
At its core, \name{} consists of three central components, as depicted in \Cref{fig:components}.
\begin{figure}
    \centering
    \includegraphics[width=\columnwidth]{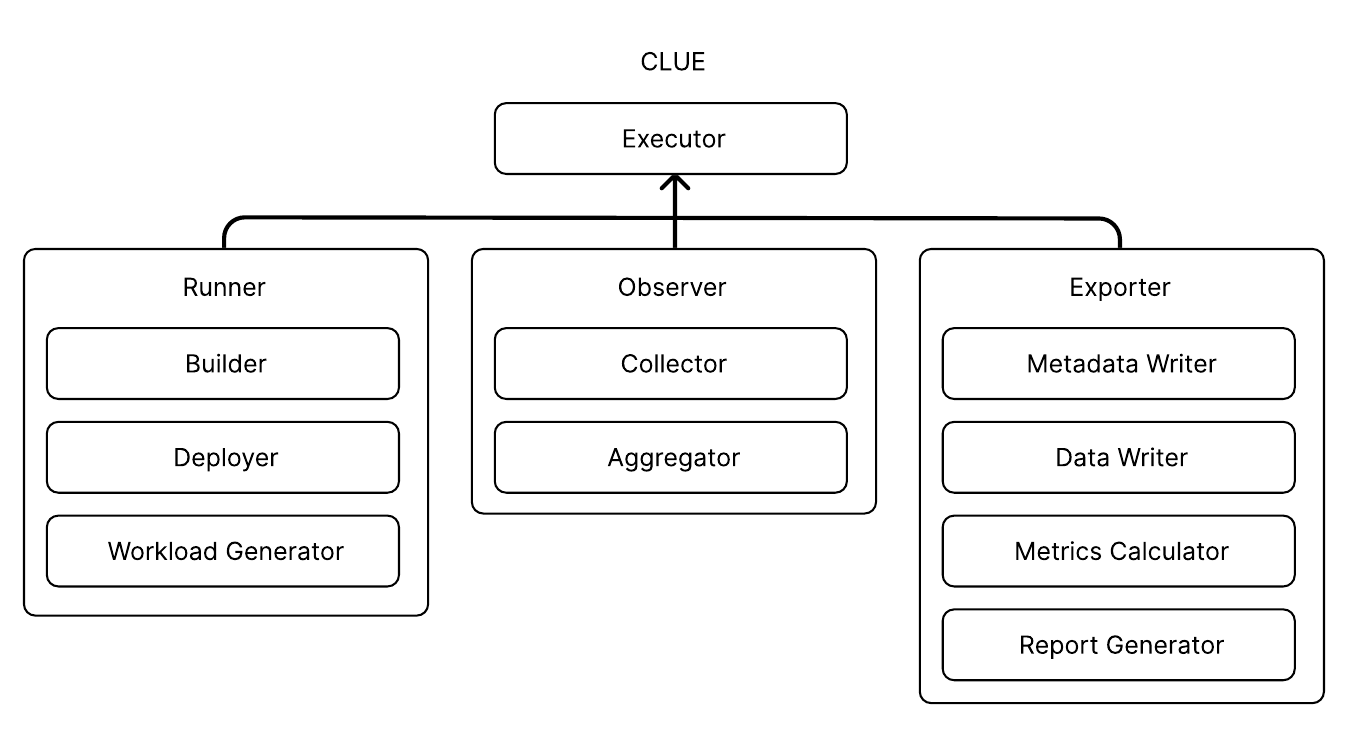}
    \caption{The high level architecture of \name{}, showing the three main components and sub-components.}\label{fig:components}
\end{figure}
Firstly, the experiment \comp{runner} is responsible for deploying the system under test (SUT), the workload generator, and any additional infrastructure services needed to observe the SUT. 
The experiment runner is separated into multiple sub-components, such as a \comp{Builder, Deployer}, and \comp{Workload Generator}. 
In its current form, we focused on Kubernetes as an execution platform, thus the \comp{Deployer} is wrapping \texttt{Helm} and the \texttt{Kubernetes API}. 
Similarly, the builder is wrapping \texttt{docker buildx}, but also allows for custom bash scripts. 
For testing and validation, we also support \texttt{kind}\footnote{\url{https://kind.sigs.k8s.io/}} to allow for local testing without the use of a cluster.
Lastly, the workload \comp{Deployer} currently supports \texttt{Locust}-based\footnote{\url{https://locust.io/}} docker files, where each experiment definition can specify which locust files should be used in a provided docker container. 
While Kubernetes is relatively limited in comparison to all possible cloud-native deployments, it already allows for the testing of several deployment approaches, such as serverless computing, e.g., through \texttt{KNative}\footnote{\url{https://knative.dev/docs/install/}}, mesh-based approaches using systems like \texttt{istio}\footnote{\url{https://istio.io/}} and multi-cloud approaches using technologies such as \texttt{kubeStellar}~\footnote{\url{https://github.com/kubestellar/kubestellar}}.
Other extensions to the \comp{Deployer} and build could be to use Terraform or CloudFormation instead to support also deployments outside of Kubernetes.

\begin{table}[]
\centering
\caption{Measurement Points for the Different Layers in \name{}}\label{tbl:measurement_points}
\renewcommand{\arraystretch}{1.4} %
\resizebox{\columnwidth}{!}{
\begin{tabular}{l l l}
\hline
\textbf{Layer}                  & \textbf{EE-Relevant Measurements}                                                                     & \textbf{EE Measurement Method}                                                  \\ \hline
\textbf{Application Layer}      & \begin{tabular}[t]{@{}l@{}} - Request count \\ - Custom metrics (useful work)\end{tabular}                & Custom instrumentation  \\
\midrule 
\textbf{Service Layer}          & \begin{tabular}[t]{@{}l@{}} - CPU utilization (\%)\\ - Memory utilization (\%)\end{tabular}             & Kubernetes metrics server                                                       \\ \midrule
\textbf{Platform Layer} 
                                & \begin{tabular}[t]{@{}l@{}}\emph{Resource consumption:}\\ - CPU utilization (\%)\\ - Memory utilization (\%)\end{tabular} & \begin{tabular}[t]{@{}l@{}}Kubernetes metrics server\\ Prometheus node exporter\end{tabular} \\
                                & \begin{tabular}[t]{@{}l@{}}\emph{Energy consumption:}\\ - Pod energy consumption (Wh)\end{tabular}         & Kepler                                                                          \\ \midrule
\textbf{Isolation Layer}        & - Host energy consumption (W)                                                                          & Scaphandre                                                                      \\
\midrule
\textbf{Physical Layer}         & - Device energy consumption (W)                                                                        & External power meters                                                           \\ \hline
\end{tabular}
}
\end{table}

Secondly, the runtime \comp{Observer} is a component that is responsible for collecting all necessary measurements needed to compile the metrics calculations and plotting. 
The observer combines a set of \comp{Collectors} that runs every few seconds to fetch measurements and an \comp{Aggregator} that combines the measurements of different collectors to build time series observations of the system under test.

These \comp{Collectors} are loosely coupled to the layers of the cloud-native application we can observe, see \cref{tbl:measurement_points}. 
We base the observations on \texttt{Prometheus} as the interface to store and fetch these measurements, where the \comp{Collectors} are responsible for enriching these measurements with other meta-data, e.g., using the Kubernetes-API.
We rely on measurement points distributed across all layers (see \Cref{fig:layers}) of the SUT. 

In the application layer, we collect request-response data from the \comp{Workload Generator}, i.e., a client perspective, that shows the performance and quality of request responses.
On the service level, we collect custom metrics, relevant to the overall application, i.e., recommendations per minute or checkouts per minute.
At the runtime level, which in the case of \name{} is mostly on the pod level, we measure the resource usage, energy consumption, and life cycle events. 

At the platform level, we measure the resource consumption, network traffic, and energy usage of Kubernetes.
For platform level energy measurements, \name{} currently supports two approaches, Kepler\footnote{\url{https://sustainable-computing.io/}} and Scaphandre\footnote{\url{https://github.com/hubblo-org/scaphandre}}. 

Both support the Intel Running Average Power Limit (RAPL) interface, while Kepler also offers support for cpu-based estimation, and Intelligent Platform Management Interface (IPMI). 
Scaphandre on the other hand also supports virtualized environments with qemu, thus, also offering a path for cloud vendors to integrate energy measurements in multi-tenant environments.
In the physical layer, we allow for external power meters to measure the total power consumption of the system. 
Specifically, in \name{} we use implemented an API for smart plugs from Tapo\footnote{\url{https://www.tp-link.com/de/home-networking/smart-plug/tapo-p115/}} which also reports all measurements into Prometheus.

Lastly, the experiment \comp{Exporter} takes the collected measurement data and computes metrics and plots for each experiment. 
It is responsible for exporting necessary experiments and deployment metadata to enable the reproduction of experiments. 
Here, \name{} uses Python pandas and seaborn libraries to generate the reports and export all metadata in JSON files.
However, we store all measurement data in CSV files, so other analysis tools can be used to process the same data.

\subsection{Usage} 

\begin{figure}[t]
    \centering
    \includegraphics[width=0.94\columnwidth]{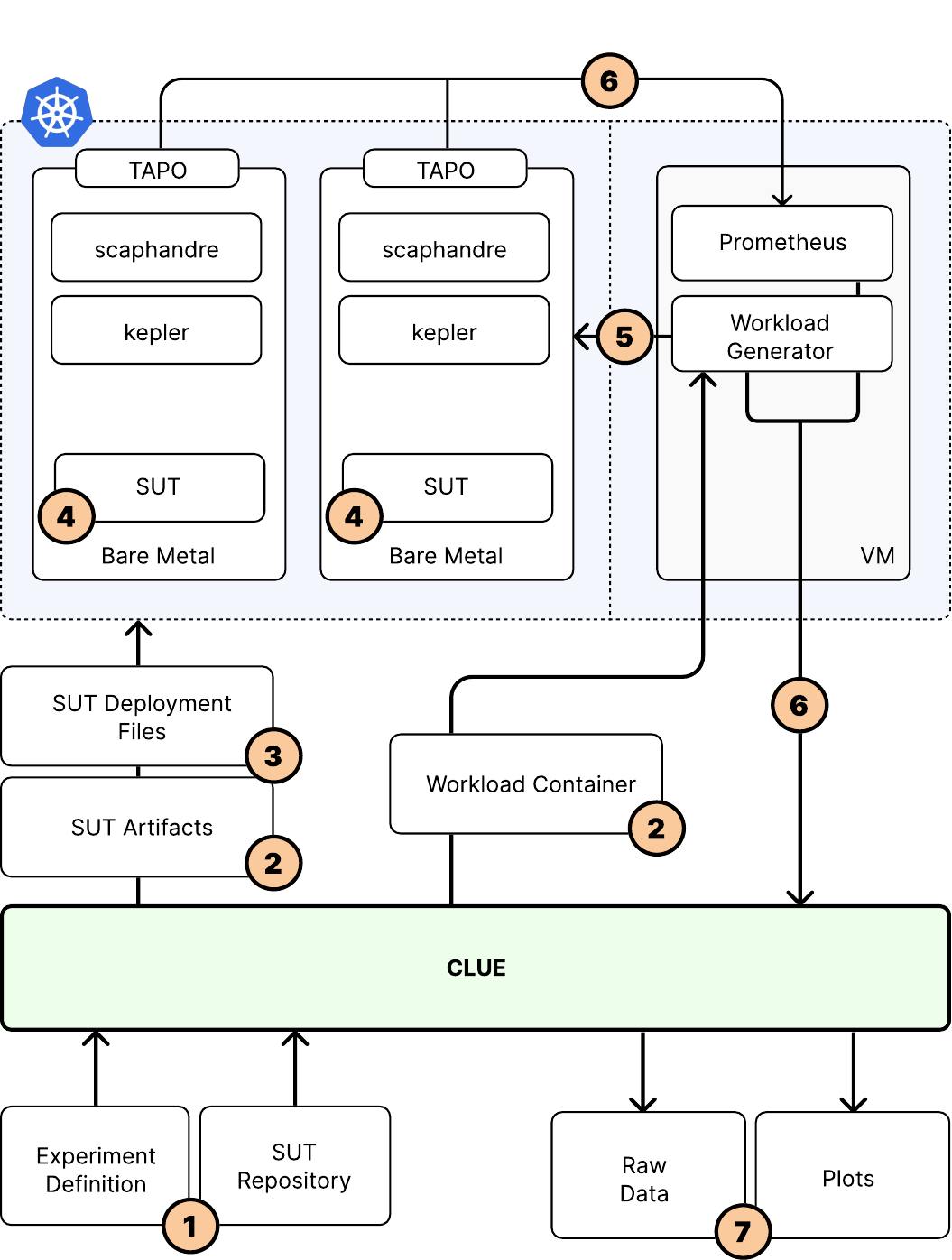}
    \caption{Steps and operations that happen in \name{} during an experiment.}\label{fig:exparch2}
\end{figure}

Using \name{} requires the user first to provide a git repository of the SUT as well as a repository for the workload generator.
We assume that every branch of the SUT repository represents a tactic of the application that needs to be evaluated. 
Lastly, the user needs to provide some final descriptions of the SUT, the intended execution environment (credentials), and additional parameters, such as the duration and repetition of experiments, see step~\circled{1} in \Cref{fig:exparch2}.

Once the user triggers an evaluation, \name{} will first build the necessary artifacts \circled{2} and either publish them to a public repository or load them into the execution environment, e.g., kind.
Following that, we take the IaC descriptions of the SUT and apply any necessary modifications, e.g., injecting information about the execution environment or the resource limits relevant for the evaluation in step~\circled{3}.

After this preparation step, \name{} deploys the SUT~\circled{4}. 
We assume that the observer and all necessary infrastructure components, e.g., Kepler, are already installed. 
We let the SUT settle before starting the workload generator in step~\circled{5}. 
Here \name{} supports two modes, either we co-locate the workload generator in the same execution environment or on the user's machine locally.
For the co-located mode, we always pick a node that is not used for the SUT to eliminate any noisy neighbor effects.

During the workload execution, \name{} uses the observer to continuously pull results \circled{6} from the execution environment. 
Moreover, we observe the SUT and workload containers directly to detect any issues until the conclusion of the experiment.
Once, the experiment is done, some additional data from the workload generator is saved before we clean up all transitionary components.

Depending on the configuration used, we terminate the SUT or repeat another workload after a settling period.
Finally, in step~\circled{7}, we compile all the raw data into a report that is provided to the user. 
Besides the report, all raw data, modified deployment files, and experiment configurations are stored in a reproduction artifact.

\section{Showcase}\label{ch:eval}
We provide a first preliminary showcase of \name{} by comparing targeted changes to the TeaStore\cite{eismann_teastore_2019} microservice in terms of sustainability and service quality.

\subsection{Evaluation Protocol} \label{sec:evalprot}

For all our evaluations, we utilized a self-hosted Kubernetes cluster (v1.28) consisting of 4x~KVM-based virtual machines and 2x~bare-metal machines, all based on x86 Linux. 
The two bare-metal machines are based on an AMD Ryzen 5 3600 (6 cores) and Intel i7-6700K (4 cores) each with 32 GB of main memory.
All VMs are backed by Xeon Silver 4114 CPUs and have also 32 GB of memory.
The VMs are receiving RAPL data from Scaphandre based on VM consumption, which is further exposed to Kubernetes through Scaphandre agents running on each VM. 
For the bare-metal machines, we use Scaphandre and Kepler directly. 
Moreover, the bare-metal machines are each plugged into a Tapo 115 smart meter that reports the total energy consumption of the machine.
All measurements are integrated into a Kubernetes-based Prometheus running in one of the VM-nodes of the cluster.
Besides the energy measurements, we also use the Kubernetes metrics server and the Prometheus node exporter to gain resource utilization and node utilization measurements.

\newcommand{\workload}[1]{\textbf{\texttt{#1}}}
For the TeaStore we implemented four different workloads, one based on a simulated day-night shaped pattern (\workload{shaped}) with two peak times around 9 AM and 5 PM with smaller peaks in between and lows at night. 
Next, we implemented a workload that terminates after 1000 requests (\workload{fixed}), and a workload that simulates a moderate amount of users with exponentially increasing breaks between active requests (\workload{pausing}).
The \workload{pausing} workload is intended to evaluate idle behavior of the changes, specifically at which idle duration the system is able to reduce its energy needs, while the fixed workload is supposed to give a direct comparison even if one of the deployments is capable of handling for more requests in the same time period. 
Lastly, we implemented a classical linear stress test that simulates up to 1000 parallel users performing interactions on the SUT (\workload{stress}), which we mainly use to see how the failure rate and resource utilization change under peak demand.

For this showcase, we implemented two adapted deployment versions of the baseline TeaStore \workload{Microservice (MS)} architecture, one as a Monolithic deployment and one as a partially Serverless deployment. We also added two variants of \workload{MS} with an improved Java runtime and a service reduction.

For the \workload{Monolith (ML)}, we merged all services other than the database into a single deployment package.
We sized the database such that it is able to handle many parallel requests. 
In the \workload{Serverless (SL)} case, we partially migrated some of the TeaStore services to KNative \footnote{KNative: \url{https://knative.dev/}}. 
We used a route-based migration approach, where each endpoint of the auth services was moved into a separate function, each of which is limited to 500~MB of Memory and 500~mCPU.
All services are using resource limits based on the recommendations of the TeaStore developers. For the monolith, each pod is allowed 2000 mCPU and 3000~MB.
For fairness, we used Java for all variants and allowed all services (including the Monolith) to scale up to 3 replicas. 
For a fourth version, we explored \workload{Service Reduction (SR)} by configuring TeaStore through Helm to not use the recommendation service.
Finally, we include a \workload{Runtime Improvement (RT)} variant, in which we swap out the Java runtime to GraalVM for OpenJDK 17 \footnote{\url{https://www.graalvm.org/release-notes/JDK_17/}}.

For each workload and variant, we ran 4 repetitions with a mean run duration of 15 minutes each. 
We performed automated data cleaning to remove outliers and faulty measurements and repeated any run that exhibited issues during the deployment or cleanup of the SUT. %

Latency and failure measurements we obtained directly from locust, from which we calculated our quality metrics in \Cref{tab:main}. 
We calculated cost by evaluating the provisioned resources and actual scaling behavior using AWS's cost model for AWS Lambda (serverless) and AWS Elastic Container Service (pods) as a reference.
We used the resource provisioning configuration to derive the resource utilization ($RU$) and the platform overhead ($RO$).
We calculate $WR$ using the total wattage reported by Kepler for the SUT divided by the number of successful requests reported by locust.
For $RE$, we determine for each second of the experiment if a pod was over-provisioned, e.g., utilizing less than 49\% for both CPU and MEM while another replica had enough resources available, in which case we added the wattage consumed by these over-provisioned pods together to calculate the \emph{scaling waste}.
\subsection{Evaluation Results}

Each of the variants implements a common architectural \emph{tactic}, targeting multiple of the layers \Name{} can measure in. We evaluate each of them by discussing the \emph{expectations} associated with that tactic in relation to the actual \emph{observations} based on what we measured compared to the baseline \workload{Microservice (MS)} variant. 
For all, we only considered the observations of the software measurement tools (i.e., Kepler, Node exporter). We used the external power only for internal validation to show that \name{} can also generate insights in a fully managed cloud environment.

\begin{table*}[]
    \centering
    \caption{Quality and performance metrics comparing three different architectures for (\texttt{pausing} - \texttt{stress}) workloads, with \low{best} and \high{worst} performance each. }
    \label{tab:quality_cost}\label{tab:main}
    \resizebox{\textwidth}{!}{
    \begin{tabular}{lcccccc}
\toprule
 & \multicolumn{2}{c}{Latency (Lat)} & Failure Rate [\%] (FR) & \multicolumn{3}{c}{Costs} \\
Feature & p50 [s] & p95 [s] &  & Total Cost [$\$$] (TC) & Consumed [$\$$] & Per Request [\textcent /1000] \\
\midrule
Microservices (Baseline) & 0.06 - 6.31 & 0.17 - 16.37 & 3.5 - 11.51 & 0.58 - 0.84 & 0.27 - 0.41 & 24.01 - 0.26 \\
Monolith & \low{0.01} - \high{23.23} & \low{0.04} - \high{42.78} & \low{0.89} - \high{41.80} & \low{0.16} - \low{0.26} & \low{0.08} - \low{0.11} & \low{10.10} - \high{0.77} \\
Serverless & \high{0.75} - 4.68 & \high{1.76} - 15.38 & \high{5.1} - 9.31 & \high{4.09} - \high{4.60} & \high{0.67} - \high{0.94} & \high{63.49} - 0.53 \\
\midrule
Runtime Replacement & 0.02 - \low{1.36} & 0.10 - 12.42 & 2.3 - \low{0.03} & 0.58 - 0.82 & 0.27 - 0.40 & 23.11 - \low{0.10} \\
Service Reduction & 0.06 - 2.61 & 0.20 - \low{8.36} & 1.9 - 1.78 & 0.69 - 0.86 & 0.28 - 0.41 & 24.98 - \low{0.10} \\
\bottomrule
\end{tabular}

    }
    
\end{table*}

\begin{figure*}
    \centering
    \includegraphics[width=\linewidth]{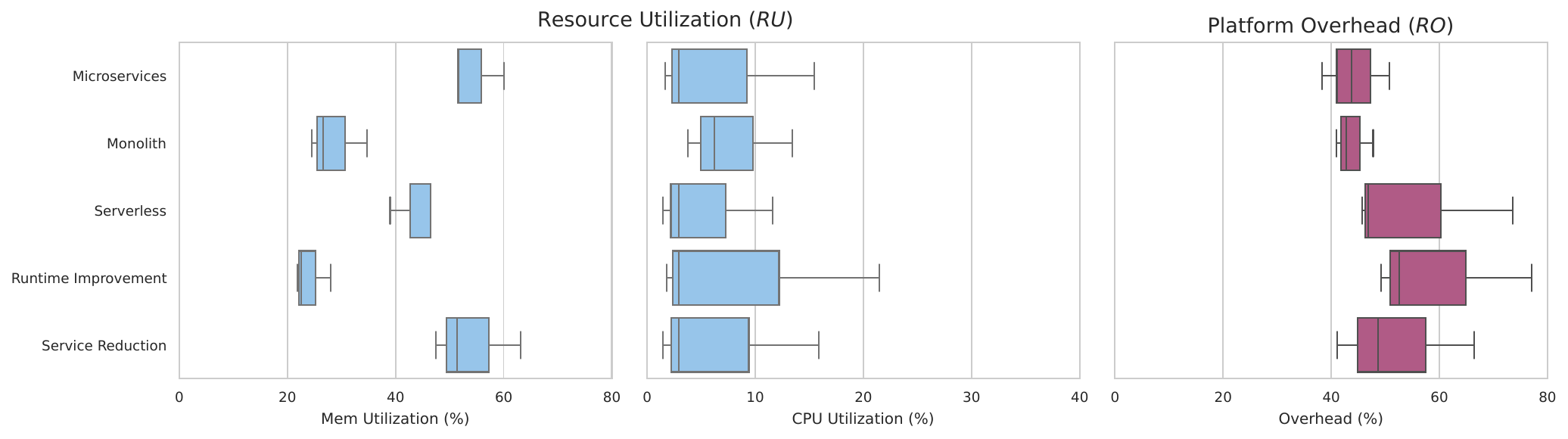} 
    \caption{Resource utilization and platform overhead for the five variants. For utilization, we compare actual consumption vs. provisioned resources. For overhead, we calculate the load not caused by SUT.}
    \label{fig:util}
\end{figure*}

\begin{figure*}
    \includegraphics[width=\linewidth]{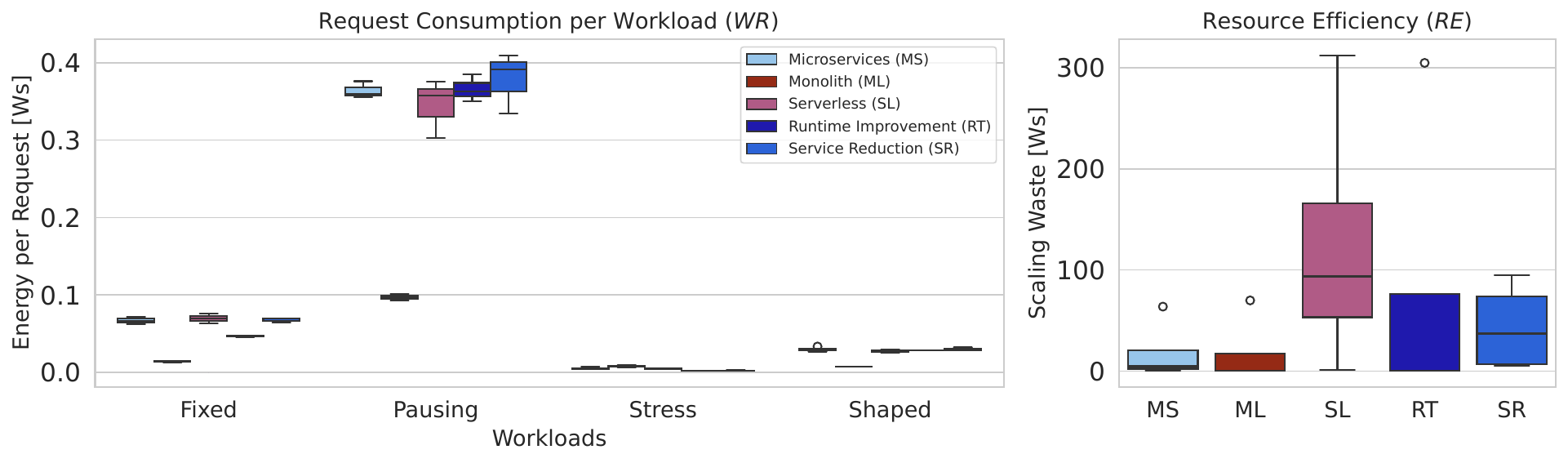}
    \caption{Consumption per requests for different workloads and wasted energy due to under-utilized scaling (average of all workloads)}\label{fig:wrre} %
\end{figure*}

In \cref{fig:wrre}, we show the energy consumption per request for different workloads, as for some variants these differ substantially.
\Cref{fig:util} shows the difference in resource utilization and the calculated platform overhead.
\Cref{tab:quality_cost} compares the key performance, QoS and cost metrics, and relates them for the three deployment variants.

\newcommand{\tactics}[0]{\textit{\newline Tactics:}}
\newcommand{\expectations}[0]{\textit{Expectations:}}
\newcommand{\results}[0]{\textit{ Observations:}}

\paragraph{\textbf{Monolith (MO)}} 
\tactics{} 
\emph{Reduce overhead by removing intermediaries and abstractions}, a ``strategy for reducing computational overhead and energy demands''~\cite{paradiskazman_energytacticsreview_2021}; Contrary to \emph{Apply Granular Scaling}, i.e. ``breaking down the workload into smaller components'' which ``results in a better match between the physical resources and the workload''~\cite{voslago_energytacticscloud_2022}

\expectations{}
On the one hand, we expected the variant to be less adaptive to changing workloads, as it is less granular.
On the other hand, we expected less overhead and thus improvements to energy efficiency due to fewer runtimes and platform components involved. We were however unclear which of the factors would dominate, and how the variant would affect Quality of Service (QoS) in our case.

\results{} 
We find that the monolithic variant does indeed not scale well: while latency slightly improves for intermediary workloads, it is significantly worse under the \workload{stress} scenario, as highlighted in the $FR$ and $Lat$ measurements of \cref{tab:main}. We also see a reduction in runtime overhead, especially, during idle time (see \cref{fig:util} and \cref{fig:wrre}). This also translates into slight increase in energy efficiency overall, although, with an unacceptable impact on QoS.

\paragraph{\textbf{Serverless (SL)}} 
\tactics{} 
\textit{Choose a fitting deployment paradigm} -- e.g., ``VMs work well with a stable, predictable workload whereas serverless architectures are suitable for bursty workloads.''~\cite{voslago_energytacticscloud_2022}, and \emph{Apply Granular Scaling}~\cite{voslago_energytacticscloud_2022} (contrary to the tactics behind the \workload{monolith} variant).

\expectations{}
Conversely to the \workload{monolith} variant, we expected the serverless parts of this variant to scale faster and more closely to the real workload, and scale to zero when not needed, improving energy efficiency while keeping baseline service quality \cite{funkelago_energytacticsexperiments_2024}. 
On the other hand, we were wary of new platform issues such as cold-starts and potential bottlenecks through other services waiting on the serverless one to scale.

\results{}
In congruence with the result for the monolithic variant, the increased overhead through e.g., significantly more runtimes and scaling events lead to a higher platform overhead (\cref{fig:util}). 
This also causes any potential scale-to-zero effects to be overshadowed by significantly higher energy consumption (see \cref{fig:wrre}) of many more deployed runtimes during typical use.
This higher energy consumption is not only caused by the added runtimes but also by the added components needed to provide serverless functionality.
Here, we saw that KNative had to deploy several energy-intensive pods to provide the serverless functionality.
Hence, in our application scenario, the expected benefits surprisingly did not materialize.

\paragraph{\textbf{Service Reduction (SR)}}

\tactics{}
\textit{Adopt use-case driven design}, especially  ``eliminating redundant software services''~\cite{voslago_energytacticscloud_2022}

\expectations{}
We expected the removal of the recommendation service to save time and energy for most of the requests, resulting in improved latency, cost, and energy efficiency at the price of service quality.
We were cautious if the configuration option of removing an entire service would yield any unexpected side effects through other services still relying on it or failing ungracefully
\cite{vitali2022towards}.

\results{}
The application no longer showed recommendations with the changed configuration. This seems to improve 
performance and failure rate overall.
However, this improvement fluctuated for different workloads. 
This potentially hints to a latent bug in TeaStore's configuration option, where other services still sometimes rely on the now missing service. Alternatively, a stronger coupling than the documentation showed between services may cause overall stability issues. The energy and cost savings were overall negligible.

\paragraph{\textbf{Runtime Improvement (RT)}}

\tactics{} 
\emph{Updating to a more performant runtime} \cite{Ournani_JVMenergyconsumtion_2021}

\expectations{}
We expected some reduction in runtime overhead due to a more efficient JIT compiler, resulting in faster and more hardware-native  execution.
We were, however, cautious of compatibility issues between the existing codebase and a new JVM leading to errors. 

\results{}
We found that the change in runtime strongly improved performance (as shown by the best observed $Lat$ in \cref{tab:main}) and lower energy consumption per request (shown in \cref{fig:wrre}). 
We did not detect the introduction of any obvious errors, with the failure rate instead improving, likely due to fewer performance bottlenecks.

\subsection{Discussion}

These showcase experiments highlight the importance of validating proposed tactics for a given architecture and environment. 
While some tactics indeed yielded expected benefits for energy consumption, others, such as the tactics in the serverless variant, did not.
This might hint at a difference in considerations in some of the tactics. 
The tactic that motivated the serverless variant (Apply Granular Scaling) might work, however, only if the system providing this fine granularity is not consuming more energy than a less granular scaling alternative.
Thus, it depends if we consider only the application's energy consumption or if we also consider the platform's energy consumption when evaluating these tactics.
Thus, we might need larger architectural patterns, that also address such side effects when building more energy-efficient applications.

Generally, we demonstrated how we could quickly compare the influence of different architectural tactics on a given service on the particular tea store stack with \Name{}.
While none of the tactics were implemented in a production-grade manner, some already yielded very strong improvements. Which, in a real scenario would allow an architect to confidently choose the next steps in furthering a more energy efficient system architecture, and also verify potential side effects.
Moreover, we could observe that variants' effects varied strongly with different workloads, highlighting the need for \Name{}'s support for custom stress scenarios as well as considering target scenarios when comparing tactics.
Notably, the variants' implementation effort varied significantly: While a change in runtime could be introduced quickly (in this case), a change in deployment paradigm required more drastic changes and careful consideration of side effects, even more so when finalizing changes for production. This adds another dimension to the trade-offs identified through \Name{} -- the (indirect) cost of applying a tactic -- when considering which one or which ones to use for a given scenario. This makes it yet more important to allow early and automatic feedback, which \name{} provides through its git integration.

While \Name{}'s primary use case is for the type of structured experimentation performed in this showcase, its component-based implementation allows observing the same metrics when deciding to fully adopt a variant in production, to see how the changes hold on production hosting and real workloads (although requiring a compatible API Gateway / proxy to measure latency and failures outside controlled experiments).
Moreover, \Name{} can be integrated into CI/CD pipelines to continuously assess the sustainability impact of ongoing development and detect any unwanted effects as early as possible.
Overall, we argue that \Name{} provides a necessary step in actively applying emerging energy efficiency tactics in practice.

\section{Conclusion}\label{ch:conclu}

\subsection{Limitations}
During the experimentation and evaluation of the data, we encountered several challenges and limitations.
Firstly, the current measurement methods for energy consumption are still very flaky. 
For example, sometimes, Kepler would report almost no consumption data, even though the external power meter showed a change in energy consumption. 
We also encountered issues due to the different CPU architectures of the two bare-metal systems, which suggests that the data used by both Kepler and Scaphandre is not entirely trustworthy. 
Here, we hope that further development for container and VM-based energy meters will improve the accuracy of results in the future. 
Moreover, while we designed \Name{} to be cloud-native, we executed the showcase experiments on a self-hosted cloud, as we wanted to have added validation from the external power meters. 
Yet, the same tools we used to report the results and that are depicted in the plots would be available on a AWS deployment as we relied only on data from the node exporter and Kepler, which work there as well. 
We would hope that developments in KVM\footnote{Integration of RAPL in VMs -- \url{https://gitlab.com/qemu-project/qemu/-/commit/0418f90809aea5b375c859e744c8e8610e9be446}} and improvements provided by cloud-vendors will allow more accurate measurements in the near future\cite{mytton2020assessing}.

Besides this, our current showcase only evaluated a single microservice architecture, based on Java.
Here, we aim to expand the evaluation to more architectures to see if the metrics and method developed for \name{} are sufficient to let practitioners evaluate existing and emerging tactics on their applications.
Lastly, the complexity of the many layers of modern cloud-native applications (\Cref{fig:layers}) creates many opportunities for errors in measurements and also presents an issue with pinpointing the source of a change in energy or resource consumption. 
For example, while we initially assumed that the scale to zero properties of the serverless deployment would show an advantage in the pausing workload, we observed the opposite. 
Due to prior experience with serverless, we were quickly able to instrument the runtime environment to pinpoint this increase to the cold starts of all authentication functions. 
However, such expertise would be needed for any particular layer to provide these insights, which might be too much of a responsibility for practitioners. 

\subsection{Summary and Discussion}
As global awareness of the environmental impact of large cloud-native applications grows, practitioners are increasingly encouraged to design their applications to be leaner and more energy-efficient, in line with the United Nations sustainable development goals. 

Until now, practitioners looking to evaluate the energy-efficiency of their application have been limited. 
While some tactics have already been evaluated in previous work \cite{funkelago_energytacticsexperiments_2024,vitali2022towards}, these experiments do not necessarily indicate that they would be effective in the context of a developer's specific application.
To address this, we developed \name{}: a comprehensive, automated, and extensible experimentation framework for developers to measure energy-efficiency-related quality trade-offs of design changes in the entire application. 
For this, we derived a set of sustainability and quality metrics that \name{} collects and offers as comparative assessments.

We showcase \name{} by comparing different promising energy efficiency tactics and apply them to the well- known TeaStore~\cite{eismann_teastore_2019} example.
First results already reveal how different scaling options of these deployments influence cost, performance, and energy consumption.
However, we also faced some limitations concerning the energy measurement and overall stability of the workloads. 
Nevertheless, the extensible nature of \name{} and the future development both from cloud vendors and energy measurement tools will widen the applicability of \name{}.
Hence, we aim to keep developing more adapters, e.g., for terraform, to enable a wider range of applications and build up a more comprehensive evaluation of applications beyond the TeaStore.

\bibliographystyle{src/IEEEtran}
\bibliography{main}
\end{document}